\documentclass[conference]{IEEEtran}
\IEEEoverridecommandlockouts

\usepackage{graphicx,
	psfrag,
	epsfig,
	epstopdf,
	amsthm,
	amssymb,
	url,
	subcaption,
	algorithm,
	balance,
	enumerate,
	color,
    url,
	setspace,
	tikz,
	pgfplots
}
\usepackage{amsmath}
\usepackage[nospace,noadjust]{cite}
\usepackage{pbox}
\usepackage{geometry}
\usepackage{multirow}
\usepackage{adjustbox}
\usepackage{array}
\usepackage{booktabs}
\usepackage{float}
\usepackage{tabularx}
\usepackage{algorithm}
\usepackage{algpseudocode}
\algnewcommand{\And}{\textbf{and}}
\usepackage{acro}

\definecolor{dark}{rgb}{0.0, 0.5, 0.0}

\newcommand{\cref}[1]{Constraint~\ref{#1}}

\newcommand{\ignore}[1]{}
\floatname{algorithm}{Procedure}
\newgeometry{left=1.2cm,right=1.2cm,bottom=1.5cm,top=1.5cm}

\DeclareAcronym{pow}{
  short=PoW,
  long=Proof-of-Work,
}

\DeclareAcronym{dlt}{
  short=DLT,
  long=Distributed Ledger Technologies,
}

\DeclareAcronym{pos}{
  short=PoS,
  long=Proof-of-Stake,
}

\DeclareAcronym{pbft}{
    short=PBFT,
    long=Practical Byzantine Fault Tolerance
}

\DeclareAcronym{pki}{
    short=PKI,
    long=Public Key Infrastructure
}

\DeclareAcronym{ca}{
    short=CA,
    long=Certificate Authority
}

\DeclareAcronym{cd}{
    short=CD,
    long=Central Directory
}

\DeclareAcronym{cms}{
    short=CMS,
    long=Certificate Management System
}

\DeclareAcronym{ps}{
    short=PS,
    long=Policy Statement
}

\DeclareAcronym{poa}{
    short=PoA,
    long=Proof-of-Authority
}

\DeclareAcronym{pob}{
    short=PoB,
    long=Proof-of-Burn
}

\DeclareAcronym{poc}{
    short=PoC,
    long=Proof-of-Capacity
}

\DeclareAcronym{sdk}{
    short=SDK,
    long=Software Development Kit
}

\DeclareAcronym{api}{
    short=API,
    long=Application Program Interface
}

\DeclareAcronym{msp}{
    short=MSP,
    long=Membership Service Provider
}

\DeclareAcronym{cli}{
    short=CLI,
    long=Command Line Interface
}

\DeclareAcronym{json}{
    short=JSON,
    long=JavaScript Object Notation
}

\begin{document}
\title{Decentralizing Trust: Consortium Blockchains and Hyperledger Fabric Explained}	

\author{\IEEEauthorblockN{Angelo Vera-Rivera}
\IEEEauthorblockA{
\textit{University of Manitoba}\\angelo.verarivera@umanitoba.ca}
\and
\IEEEauthorblockN{Ekram Hossain}
\IEEEauthorblockA{\textit{University of Manitoba}\\ekram.hossain@umanitoba.ca}
}

\maketitle

\begin{abstract}
Trust models are essential components of networks of any nature, as they refer to confidence frameworks to evaluate and verify if their participants act reliably and fairly. They are necessary to any social, organizational, or computer network model to ensure truthful interactions, data integrity, and overall system resilience. Trust models can be centralized or distributed, each providing a good fair of benefits and challenges. Blockchain is a special case of distributed trust models that utilize advanced cryptographic techniques and decentralized consensus mechanisms to enforce confidence among participants within a network. In this piece, we provide an overview of blockchain networks from the trust model perspective, with a special focus on the Hyperledger Fabric framework, a widespread blockchain implementation with a consortium architecture. We explore Fabric in detail, including its trust model, components, overall architecture, and a general implementation blueprint for the platform. We intend to offer readers with technical backgrounds but not necessarily experts in the blockchain field a friendly review of these topics to spark their curiosity to continue expanding their knowledge on these increasingly popular technologies. 
\end{abstract}

\section{Introduction}
\begin{figure*}
    \centering
    \includegraphics[width=0.8\textwidth]{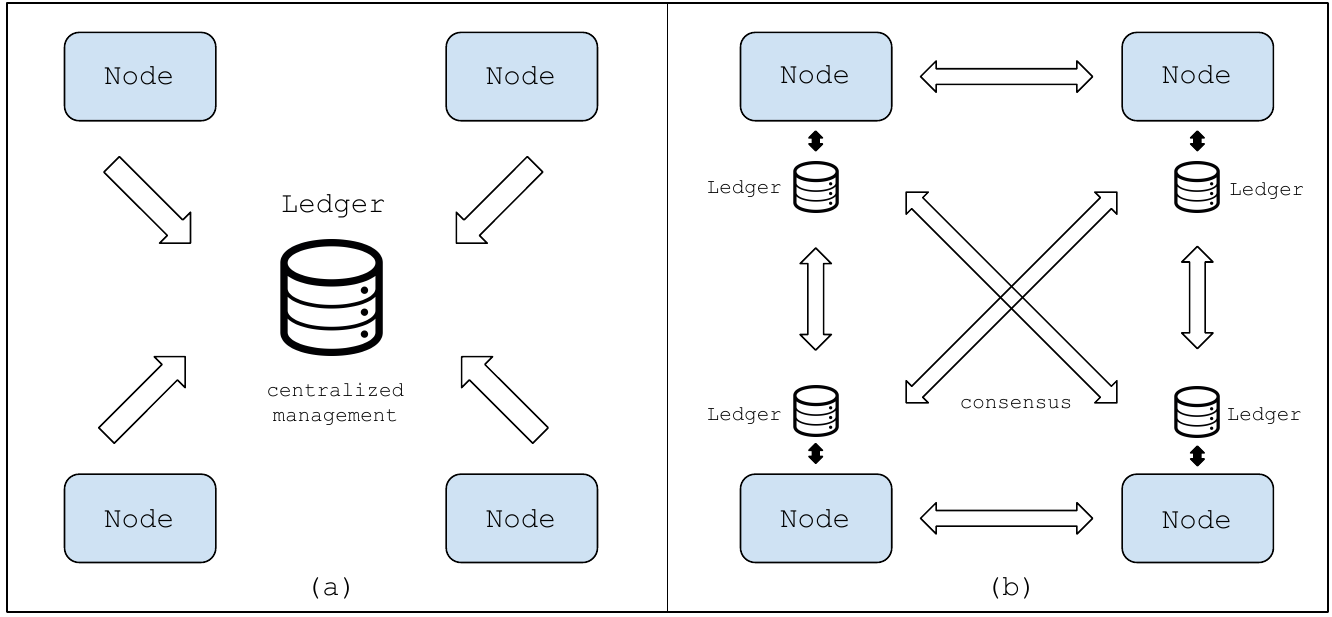}
    \caption{(a) Centralized ledger and (b) DLT}
    \label{fig:DLT}
\end{figure*}
Trust refers to the belief or confidence in an entity, such as a person, organization, or system wihitn a network, to act reliably and fairly. It is typically understood in terms of three key aspects: reliance, expectation of integrity, and predictability. Trust involves trusting individual entities or groups to meet basic safety expectations, based on the assumption that they will always act in good faith and with honesty. Building trust generally requires time ans is influenced by consistent behavior in the environment. It is often situational, as the definition of reliability and fairness can vary depending on the context and circumstances of the network. A decent trust model ensures that nodes within a network can interact reliably and truthfully with one another, which enhances confidence in the network's integrity and security policies. Moreover, trust is essential for protecting networks against security breaches, malicious attacks, and spontaneous failures. \\

Trust models are conceptual frameworks that define how trust is established, managed, and maintained within a network. In centralized trust models, systems rely on a central entity or node with sufficient authority to establish trust among all network participants. This central authority facilitates communication and manages the relationships between participants. Networks using centralized trust models are simpler to manage because they reduce the need for complex peer-to-peer negotiations, which are handled exclusively by the trusted central party. On the downside, a trusted central party also represents a liability for the network. If the central node gets compromised or become unavailable, the entire system's trust could be in danger. Additionally, the central authority model has limited feasibility in practical applications, especially for consortium architectures, where trust must be shared among multiple organizations\footnote{In technical domains, a consortium refers to collaborative frameworks where independent entities (e.g., organizations) come together to pool their resources to achieve shared objectives by adhering to mutually agreed rules and governance structures.}. \\ 

On the other hand, distributed trust refers to network models in which trust does not depend on a central authority but is instead shared among multiple participants within the network. Depending on how it is implemented, distributed trust can enhance transparency, resilience, and collaboration in a network by ensuring that measures of reliability, integrity, and predictability are openly accessed and managed collectively by all (or many) network participants. This model is based on the core principle that no single entity within the network is completely reliable for making all decisions. Instead, trustworthiness becomes a collective effort, each participant contributing to this labor. Decisions in the network are then achieved collectively through consensus mechanisms, allowing for a democratic approach to governance. Additionally, distributed trust models ensure that decision logs and data are tamper-proof and transparent, giving all participants reliable access to all the information regarding interactions within the network. Among other benefits, these models solve single-point-of-failure problems, making networks more robust and resistant to failures. \\

Blockchain systems are a specific type of distributed trust models where no single entity within the network is automatically trusted. Instead, every interaction within a blockchain network must be verified and approved independently by network participants using advanced cryptographic techniques and decentralized consensus algorithms. Also, network interactions are recorded in a distributed database known as ledger, which all participants are responsible for maintaining. This setup ensures that network information is distributed, eliminating the dependency on a single node to hold the system's records. Blockchains run on three fundamental principles: (i) the network ledger must be immutable, (ii) the network records and data must be open and transparent, and (iii) the data must be consistent and unified across all network nodes. Consortium blockchains refer to a specific blockchain framework in which multiple independent entities, such as organizations, institutions, companies, or computer nodes, collaborate to achieve shared objectives by pooling resources, knowledge, or infrastructure. These entities maintain their individual autonomy while adhering to mutually agreed rules, protocols, and governance structures. Typically, trust in consortium architecture enforces the following core ideas: collaborative governance, distributed authority, access control, accountability, and mutual benefit through a modular architecture and network services.\\

This article presents an overview of blockchain technologies, emphasizing their ability to establish trust within a network. The article focuses on the Hyperledger Fabric framework, a consortium-based blockchain platform. The draft is organized as follows. Section II presents a general overview of blockchain, discussing its key features, components, and types. It includes a comparison of consortium blockchains with public and private blockchains. Section III discusses the Hyperledger Fabric framework in detail, highlighting its trust model, components, and overall architecture. Section IV outlines a blueprint for implementing Fabric blockchain networks. This section covers network setup, deployment of Fabric components, creation of blockchain channels, deployment of smart contracts, and integration of external client applications with the blockchain network. We wrap this piece with a short conclusion section and a list of relevant references. 
\section{Blockchain Overview}
\begin{table*}
\centering
\resizebox{0.6\linewidth}{!}{
\begin{tabular}{ | p{3cm} | p{4cm} | p{5.5cm} |} 
\hline
\textbf{Level 1} & \textbf{Level 2} & \textbf{Level 3} \\
\hline
& & \\
1. Membership & 1.1 Access control & 1.1.1 Public un-permissioned\\
& & 1.1.2 Private permissioned\\
& & 1.1.3 Consortium permissioned\\ 
& 1.2 Identity Management & 1.2.1 Anonymous \\
& &  1.2.2 Public Key Infrastructure (PKI) \\ 
& & \\
 \hline
 & & \\
2. Consensus & 2.1 Governance & 2.1.1 Decentralized\\
 & & 2.1.2 Hierarchical \\
 & & 2.1.3 Centralized  \\
 & 2.2 Type of consensus & 2.2.1 \ac{pow} \\
 & & 2.2.2 \ac{pos}  \\
 & & 2.2.3 \ac{poa} \\
& & 2.2.4 \ac{pob} \\
& & 2.2.5 \ac{poc} \\
& & 2.2.6 Endorsement based \\
& & 2.2.7 Hybrid \\
& & \\
\hline
 & & \\
3. Ledger & 3.1 Permissions & 3.1.1 Open read/write\\
& & 3.1.2 Identity-based read/write\\
 & 3.2 Type of database & 3.2.1 LevelDB\\
 & & 3.2.2 CouchDB\\
 & & \\
\hline
 & & \\
4. Smart contract & 4.1 Logic structure & 4.1.1 Hard-coded logic \\
 & & 4.1.2 Programmable logic \\
 & 4.2 Coding Language: & 4.2.1 Single dedicated syntax\\
 & & 4.2.2 General purpose syntax \\
 & 4.3 License: & 4.3.1 Open source \\
 & & 4.3.2 Closed source \\
  & &  \\
\hline
\end{tabular}}
\caption{Three levels of blockchain taxonomy}
\label{table:blockchainTaxonomy}
\end{table*}
Blockchain platforms are peer-to-peer computer networks that utilize (i) sophisticated cryptography and (ii) consensus mechanisms to collectively verify and approve interactions among network nodes that do not necessarily trust one another. Blockchains store the records of network interactions in distributed databases securely maintained by all the network participants. The combination of cryptography and consensus permits the nodes to interact honestly without a central authority within the system imposing the rules. When used appropriately and in the right context, the decentralization services offered by blockchain technologies enable self-governed, self-regulated, intelligent, and secure network functions. A brief historical note: the blockchain technology was publicly introduced by Satoshi Nakamoto in his 2008 white paper "Bitcoin: A Peer-to-Peer Electronic Cash System" \cite{01}. In the paper, Nakamoto proposed the Bitcoin platform, a distributed payment system operating on a peer-to-peer network that he named blockchain. In Nakamoto's blockchain, the interactions among nodes are validated and approved through a stochastic consensus mechanism known as \ac{pow} and the network records are stored in a distributed database using a cryptographic-aided data structure maintained by all network nodes. The system was the first known implementation of a decentralized network in which participants could securely transact with one another without relying on a central node overseeing network interactions. \\

Blockchain nodes can engage with the network through transactions regulated by a collective consensus. Then, consensus can be seen as a form of "democracy" within a network. It enables the system to make collective decisions, allowing nodes to "vote" on the validity of transactions. Transactions can be hard-coded into the network, or they can be programmable. Programmable transactions are commonly referred to as Smart Contracts. The level of democracy in a consensus mechanism can vary, and some democracy levels require computationally intensive algorithms. Transactions voted as valid are grouped into data blocks connected chronologically using cryptographic methods, hash pointers specifically, creating a data structure that resembles a chain of blocks or blockchain. Data blocks with valid transactions are added to the blockchain database, also known as the blockchain ledger or simply the ledger. The ledger stores records of every valid interaction in the network, making the complete record collection the official history of the network. By maintaining a local copy of the ledger, blockchain nodes act as guardians of the historical record, promoting transparency and accountability, which is essential to establish trust in distributed systems. As a general rule, blockchain networks uphold three fundamental characteristics: (i) immutability of the ledger, (ii) transparency of information, and (iii) consistency of data across nodes. Blockchain systems derive directly from \ac{dlt}, a database framework that defines a set of cryptographic protocols to store transaction records in a distributed fashion. \ac{dlt} provides a way to access, validate, update, and manage decentralized databases. These protocols ensure that the records remain immutable, consistent, available, and transparent across the network. An illustration of DLT versus centralized systems is shown in Figure \ref{fig:DLT}. 
\subsection{Public vs. Private vs Consortium Blockchains}
\begin{table*}
\centering
\resizebox{0.7\linewidth}{!}{
\begin{tabular}{ | p{3cm} | p{2.5cm} | p{2.5cm} | p{2.5cm} |} 
\hline
 & \textbf{Public} & \textbf{Private} & \textbf{Consortium} \\ 
\hline
& & & \\
\textbf{Decentralization:} & & &  \\
Social analogy  & Libertarianism & Dictatorship & Federalism \\
 & & & \\
\hline
& & & \\
\multicolumn{1}{|l|}{\textbf{Access control:}} & & & \\
Access Permission & Unpermissioned & Permissioned & Permissioned \\
Membership service & No & Yes & Yes\\
Ranked members & No & Yes & Yes \\
 & & & \\
\hline
& & & \\
\textbf{Consensus:} & & &  \\
Governance & Distributed & Centralized & Federated\\
Type of consensus & Proof-based & Voting-based & Voting-based \\
Computation demand & High & Low & Low \\
Energy demand & High & Low & Low\\
Tx throughput & Low & High & High\\
Tx latency & High & Low & Low\\
Scalability & Low & High & High \\
 & & & \\
\hline
& & & \\
\textbf{Transparency:} & & &  \\
Data & Public & Private & Private\\
Tx visibility & Visible & Partially visible & Partially visible \\
Participant's identity & Anonymous & Disclosed & Disclosed\\
 & & & \\
\hline
\end{tabular}
}
\caption{\centering Public vs. Private vs. Consortium Blockchains}
\label{table:publicPrivateConsortium}
\end{table*}
Blockchain networks can be classified in various ways based on several factors, including type of governance, consensus, rewarding system, code-base design, identity management, etc. We can establish a taxonomy for blockchain systems based on the following blockchain services: (i) membership, (ii) consensus, (iii) ledger, (iv) and smart contract types. For more details on blockchain taxonomy, please refer to Table~\ref{table:blockchainTaxonomy}. Additionally, blockchain systems can be further segmented into three main groups based on membership and access control types. According to this segmentation, blockchains can be grouped as (i) Public, (ii) Private, and (iii) Consortium platforms. \\

Public blockchains are open platforms without restrictions on accessing the blockchain infrastructure. Anyone with an internet connection can join the network and become an authorized node. Typically, participants in public blockchains have transparent access to data, can initiate transactions, and participate in the transaction verification and block creation processes. These blockchains generally feature open-source designs and employ consensus mechanisms that prioritize high levels of decentralization. \ac{pow} and \ac{pos} are examples of consensus algorithms used in public blockchains. A key advantage of these platforms is that they operate independently of the control of any single owner. For instance, if the founding node exits the system or crashes, the network will rely on the remaining nodes to continue operating normally without disruptions. On the downside, the performance of public blockchains tends to be slow due to the use of highly democratic and decentralized consensus mechanisms with extensive computation requirements. This is a significant drawback for the scalability of these blockchains, representing a major drawback for the systems. Generally, public blockchain platforms are well-suited for applications requiring fully decentralized, transparent, and notarized trust models that promote open collaboration among network participants. Decentralized payment systems and cryptocurrencies are common applications of these blockchains. \\

Private blockchains are permissioned platforms typically governed by a single node within the network. Although they function as peer-to-peer networks, the governing node assigns specific roles to participants, normally limiting their rights regarding access to network resources and defining their acceptable behavior within the network. Generally, private blockchains are invite-only systems employing a membership service to enforce access control. Their hierarchical, role-based structure typically allows for higher transaction throughput than public blockchains. However, their highly centralized consensus can be counterintuitive to the fundamental blockchain conception, limiting their applications and use cases. Additionally, private blockchains are often closed-source, proprietary, and not subject to third-party audits. These blockchains are suitable for applications requiring cryptographic and access control services, where a central governing node can be fully trusted, and transparency is not necessarily a priority. \\

Consortium blockchains are platforms that combine features of public and private blockchains. In these blockchains, the nodes are controlled by several organizations, possibly competitors, with distinct operational and economic incentives (e.g., different telecommunication service providers operating in the same geographical area). Using consortium blockchains, these organizations could collaborate in decentralized platforms with a role-based structure, sharing control over access, permissions, security, and resources within the network. Governance in consortium blockchains is distributed among the participating organizations, marking a profound shift from centralized management in private blockchains. Then, consortium consensus protocols are managed by predefined trusted nodes from each organization, which validate transactions and create new blocks. The consensus nodes ensure the data is consistent across the network through fault-tolerant algorithms that are typically more computationally and energy efficient than those used in public blockchains. The design of consortium blockchains is generally open source with modular architectures, allowing for plug-and-play components and services that can quickly adapt to different needs and solution types.\\

To compare public, private, and consortium blockchains, we can examine four key blockchain components: (i) access control, (ii) consensus, (iii) transparency, and (iv) decentralization. A summary of this comparison can be found in Table~\ref{table:publicPrivateConsortium}.\\

\subsubsection{Access Control}Public blockchains are unpermissioned, meaning anyone with an Internet connection can join these platforms freely without any access control mechanism supervised by users with higher rank. In contrast, private and consortium blockchains are permissioned, meaning participants may need to disclose their identities to access the platform. These networks typicallu include a membership control module that verifies identities and accepts or rejects access requests. \\

\subsubsection{Consensus} Typically, public blockchains utilize highly democratic consensus algorithms, such as \ac{pow}, in which each node has voting power. However, high levels of democracy often result in longer decision-making times, low transaction throughput, and significant CPU and energy consumption. In contrast, private and consortium blockchains normally use less democratic consensus schemes. In these systems, decision-making power is consolidated among certain trusted nodes within the network. Examples of such methods include \ac{pos} and \ac{pbft}. With these methods, consensus is typically achieved by a small group of trusted nodes with veto power, often pre-designated by the network participants. The consensus schemes used in private and consortium blockchains tend to be computationally inexpensive, leading to faster decision-making times and lower energy consumption.\\

\subsubsection{Transparency} In blockchain networks, transparency can be understood in two dimensions: the visibility of data and the visibility of participant identities within the network. In public blockchains, transaction data is visible to all nodes, providing complete identity anonymity for the platform's participants. In private and consortium blockchains, transaction data is only visible to nodes if granted access. Participants are typically not anonymous since identities may need to be partially or fully revealed to access the network. \\

\subsubsection{Decentralization} Public blockchains exhibit high levels of decentralization, meaning there is likely no concentration of power within the network. All nodes have equal rank, rights, and freedoms in these systems, resembling a libertarian utopia. In contrast, private blockchains are typically centralized, often governed by a single leader who dictates rules, creating a form of digital dictatorship. Consortium blockchains, on the other hand, adopt a division of power approach. This model shares core functions such as identity management, transaction verification, and block generation among a consortium of participant organizations. This structure reflects a federated digital democracy, balancing the need for a robust trust model with a yet scalable architecture. 
\section{The Hyperledger Fabric Framework}
\begin{figure*}[H]
    \centering
    \includegraphics[width=0.8\textwidth]{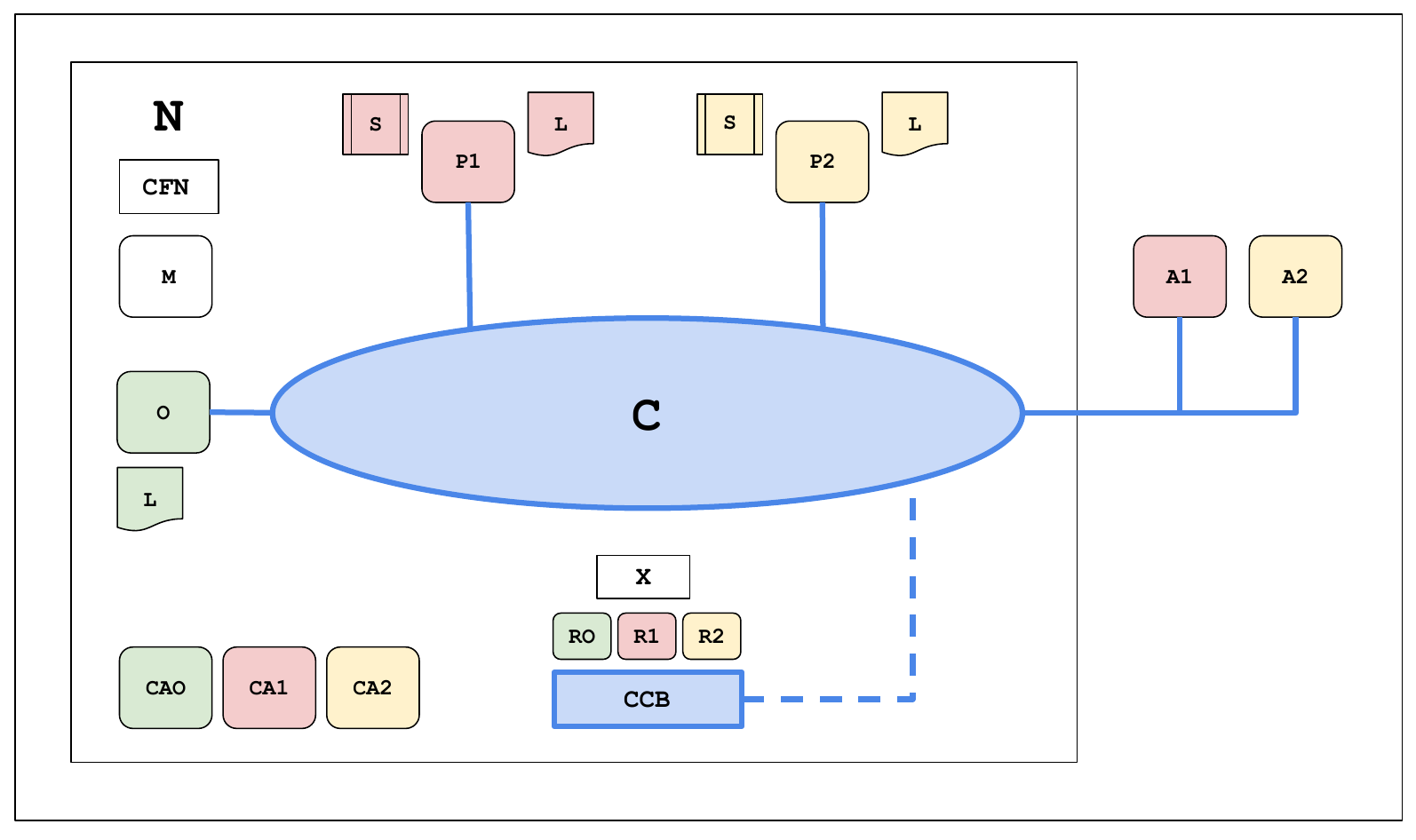}
    \caption{Hyperledger Fabric trust model}
    \label{fig:trustModel}
\end{figure*}
\begin{figure*}
    \centering
    \includegraphics[width=0.8\linewidth]{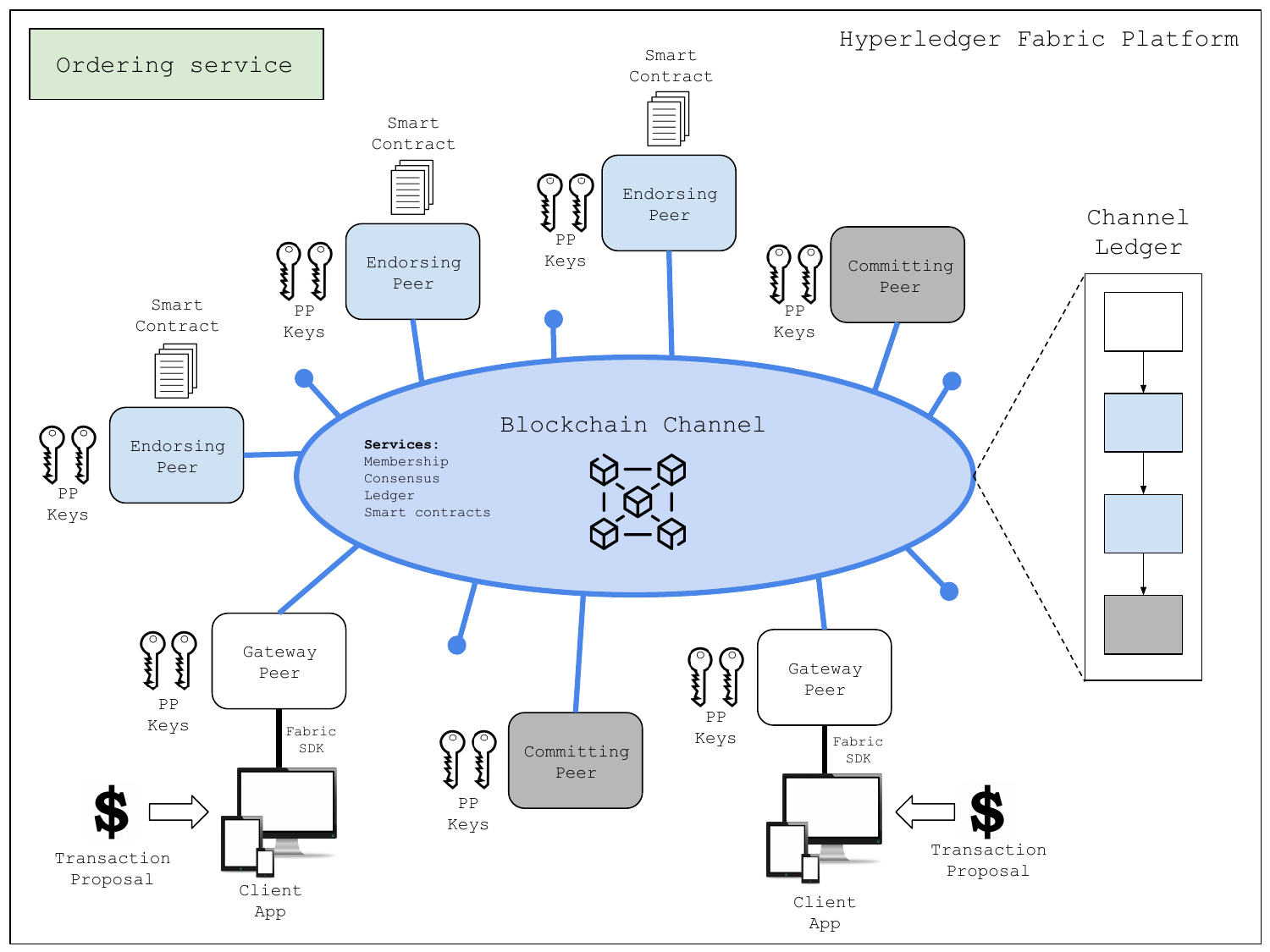}
    \caption{Fabric trust model in action.}
    \label{fig:FabricTrustModel}
\end{figure*}
In 2015, the Linux Foundation launched the Hyperledger Project~\cite{02} to develop versatile blockchain platforms with flexible frameworks that can adapt to decentralized solutions across various industries. Rather than promoting a single blockchain standard, the Hyperledger Project supports a community of software developers and industry leaders collaborating on developing blockchain technologies using an open-source approach, modular architectures, and flexible intellectual property rights. This encourages industries to adopt the technology gradually. Hyperledger Fabric is one of the blockchain platforms within the Hyperledger Project. It features a general-purpose design with a consortium model, offering lightweight, fast, scalable, and secure blockchain services~\cite{03}. Hyperledger Fabric has gained significant traction across various economic sectors, including healthcare, finance, supply chain, and retail commerce.\\

Hyperledger Fabric is a permissioned blockchain~\cite{04} designed to provide membership, consensus, ledger, and smart contract services, enabling the creation of versatile decentralized solutions. Fabric's technical design is characterized by four key features: (i) an open-source approach, (II) a modular architecture, (iii) identity-based roles, and (iv) flexible programmable logic. First, Hyperledger Fabric is open-source, allowing users to utilize, modify, and distribute the blockchain code without restrictions. This has encouraged a community of over 35 partner companies and thousands of developers worldwide who work together to develop blockchain applications for numerous industry verticals. Second, Fabric's modular architecture enables configurable membership, consensus, ledger, and smart contracts, which are flexible to various business needs. Third, Fabric incorporates identity-based roles. Fabric nodes are assigned digital identities directly linked to permissions to handle resources, access data, and allow behavior within the network. Finally, Fabric supports flexible smart contracts, which can be programmed using general-purpose languages such as \textit{Java, Go, and Node.js}. This contrasts with other blockchain platforms like Ethereum, which require platform-specific programming languages (e.g., Ethereum Solidity) for smart contract development. 
\subsection{Fabric Trust Model}
To explain the core services offered by the Hyperledger Fabric model, we have divided them into four main components: (i) membership, (ii) consensus, (iii) ledger, and (iv) smart contracts. A detailed explanation of the components will be provided next. An illustration of Fabric's trust model can be found in Figure \ref{fig:FabricTrustModel}. \\ 
\subsubsection{Membership}
Hyperledger Fabric is a permissioned blockchain, meaning participant nodes must have a verified identity. Each network component, including organizations, peers, and client applications, is assigned a unique identity. Fabric uses a \ac{pki} framework to create, maintain, store, and revoke these cryptographic identities. A \ac{pki} consists of hardware, software, and cryptographic infrastructure that assigns a pair of public and private keys to network participants through an enrollment process that generates digital certificates. The most commonly used format for these digital certificates in \ac{pki} systems is the X.509 standard. \ac{pki}s provide essential trust services that enable secure communication over the Internet. These services are widely utilized in various applications, including online e-commerce, banking, and confidential messaging. Trust services are built on three key attributes: (i) confidentiality, (ii) integrity, and (iii) authenticity.  Confidentiality ensures that external parties cannot access data exchanged between two nodes during peer-to-peer communication. Integrity guarantees that data within the \ac{pki} network is tamper-proof, meaning that it becomes tamper-evident if the data is compromised. Authenticity confirms that the nodes involved in peer-to-peer communication are legitimate. This process utilizes a digital certificate and a private key to identify nodes and validate the transferred data. \\

The PKI framework consists of four essential components: (i) \ac{ca}, (ii) \ac{cd}, (iii) \ac{cms}, and (iv) \ac{ps}. First, \ac{ca}s are trusted entities within the \ac{pki} infrastructure responsible for approving, issuing, and publishing public key and digital certificates for network nodes. They use public key encryption-decryption algorithms (e.g., RSA) to sign digital certificates with their public-private key pair. This encryption process facilitates secure communication of private information over a network, using publicly available information and a scheme for cryptographic signature verification. Second, \ac{cd}s are the secured physical locations where key pairs are stored and managed. Third, the \ac{cms} is responsible for delivering the digital certificates issued by the \ac{ca}s to their intended recipients. Finally, the \ac{ps} is an official document issued by the \ac{pki} that is accessible to the public. It outlines the requirements, operational details, and lifecycle of the \ac{pki} services. The Hyperledger Fabric framework uses a \ac{pki}-based membership service to verify participants' identities within the blockchain network. Participants, including administrators, peers, and client applications, must have a public-private key pair issued by a trusted \ac{ca} to sign transactions submitted to the network. This capability allows Fabric to enforce identity-based permissions over resources, access to data, and permitted behavior within the network. The permissioned nature of Fabric makes it appealing for application scenarios where security and privacy are essential concerns.\\ 

\subsubsection{Consensus}
The core idea of \ac{dlt} is that records must be kept distributed among the participant nodes within the network. For this purpose, it is essential to implement mechanisms that verify and organize the records added to the ledger. A consensus mechanism is a transaction validation algorithm designed to establish agreement on the data circulating in an unreliable network. It ensures that data flows are monitored, transaction blocks are constructed, and network data is safeguarded against security threats. The most well-known consensus mechanism to date is the \ac{pow} algorithm. \\

In permissionless blockchains like Bitcoin, any node can participate in the consensus mechanism. The nodes can verify, order, and package transactions into blocks, which are then proposed to the network. The consensus in these blockchains relies on probabilistic algorithms to ensure data consistency to a certain level of confidence. However, these probabilistic algorithms may be susceptible to ledger forks, where some nodes receive blocks with transactions arranged differently, leading to varying interpretations of the network's history. In contrast, Hyperledger Fabric operates as a permissioned blockchain with a slightly different structure. In Fabric networks, consensus involves more than just agreeing on the data circulating within the network. Specifically, Fabric's consensus plays a critical role throughout the entire lifecycle of transactions, from proposals and endorsements to ordering and committing events. Furthermore, Fabric's consensus can be understood at two distinct levels: (i) consensus at the peer level and (2) consensus at the ordering service level. \\

\begin{figure*}
    \centering
    \includegraphics[width=0.7\linewidth]{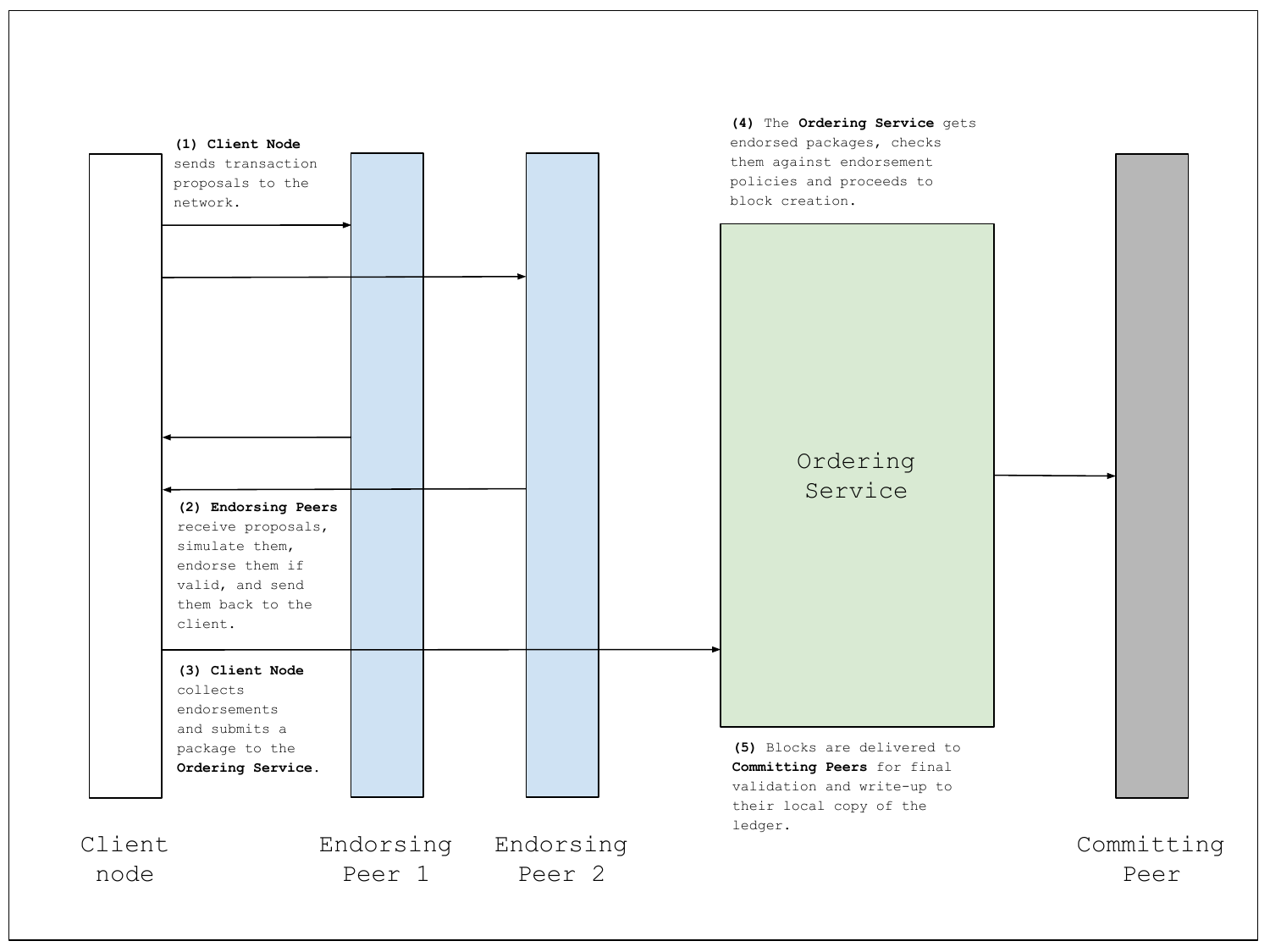}
    \caption{Transaction lifecycle within the Hyperledger Fabric platform}
    \label{fig:transactionLifecycle}
\end{figure*}

Consensus at the peer level relies on endorsements~\cite{05}, which are check and balance validations that occur during the transaction lifecycle before transaction proposals reach the ordering service. Specifically, an endorsement is a validation made by endorsing peers, who are special Fabric nodes granted specific privileges to evaluate the validity of transactions. The endorsement process involves simulating a transaction proposal using the copy of the chaincode hosted on the endorsing peer. If the simulation produce an output that matches the proposal, the endorsing peers will sign the proposal and return the signed transaction to the client that originated the transaction. Endorsement policies are outlined in the network's configuration files, where the administrator designates trusted endorsing peers responsible for simulating and signing transactions on the blockchain. The role of endorsements within the transaction lifecycle is illustrated in Figure~\ref{fig:transactionLifecycle}. \\

Consensus at the ordering service level begins with the verification of transaction endorsements. Once transactions are signed by endorsing peers and returned to invoking clients, those clients prepare a package to be submitted to the ordering service. This package includes the transaction results along with the collected endorsements. The endorsements are validated against the endorsement policies defined within the network (e.g., AND, OR, nOutOf). Suppose an endorsed transaction complies with the policies. In that case, it is declared valid, and the consensus process continues with ordering transactions, creating blocks, and broadcasting new blocks to committing peers. The ordering service is responsible for both ordering transactions and creating blocks. While the ordering service component may initially appear centralized, it is actually made up of a set of ordering peers belonging to different participant organizations. Within the ordering service, the ordering peers operate under a leader-follower structure using the Raft consensus algorithm. The reliability of the ordering service depends on Raft's crash-tolerant nature. In summary, the consensus model in Fabric networks involves multi-stage checks and transaction verifications to ensure data reliability within the blockchain network.\\
\subsubsection{Ledger}
The term "ledger" refers to a financial book or system that records transactions in a specific order. In the context of Fabric, the ledger serves as the physical database that maintains a permanent and indisputable history of blockchain records, storing facts and data about the network. While the facts and data of the network, known in Fabric as "the world state," change with the execution of transactions, the complete records of those transactions remain unique and immutable. In Fabric networks, the ledger is accessible on a blockchain channel basis, meaning only participating peers in a channel can access the channel's ledger. This design choice is intentional, as it helps maintain the privacy of transaction records within a channel. Ledgers in Fabric networks consist of two components: (i) the world-state database and (ii) the blockchain database. \\

The world state database contains the current states of objects within the Fabric network, represented as key-value pairs. The ordering service updates these states each time a transaction completes its lifecycle. The idea of a world state is beneficial because it allows programs to query the current states of objects without looping through the network's entire transaction history. Client applications can submit transaction proposals to query, update, and delete states in the world state. However, only accepted transactions can make changes to the database. Depending on the configuration, the world state provides advanced operators for efficient querying and storage of state values. Whenever a new peer is added to the network, a local instance of the world state is created for that peer. If a peer crashes, the world state can be regenerated from the network's transaction records when the peer restarts. \\ 

The blockchain database keeps a comprehensive historical record of all valid and invalid transactions within the Fabric network. It specifically tracks how Fabric objects reach their current states. Transactions in the blockchain database are organized into data blocks linked in chronological order, forming a blockchain structure. Each block's header contains the cryptographic hash of the transactions contained in the block, as well as the hash of the previous block's header. This cryptographic linkage ensures that the blocks are securely connected, preventing any modification or tampering of their data. Fabric utilizes the SHA256 hash function for these hash operations. The blockchain database is implemented as a file, whereas the world state is maintained in a NoSQL database. \\

\subsubsection{Smart Contracts} 
Smart contracts are immutable, self-executing software programs available on a blockchain that define objects within the network and provide instructions for modifying them. Before they can be used, blockchain participants must install, store, and approve smart contracts. They play a crucial role in enforcing the business logic of a blockchain application, determining how data and transactions are represented within the network. Smart contracts automate the execution of agreements pre-approved by participant organizations, eliminating the need for a central authority overseeing the enforcement of these agreements. Smart contracts are, in fact, the executable logic of a blockchain application, using transaction functionality to update facts and data in the network. \\

Smart contracts play a crucial role in establishing the terms, definitions, and processes governing interactions among Fabric network participants. With these contracts, Fabric clients and peers can submit transaction proposals to query, update, or delete the state of an object within the network. The blockchain database records these transactions to verify their validity. However, the world state database is only updated if the transactions are valid. Smart contracts come with configurable endorsement policies that specify which organizations and peers within the network must approve (i.e., endorse) proposals for them to be considered valid. This endorsement scheme in Fabric is distinct from that of other blockchains, where any node can issue endorsements. Fabric allows configurable endorsement policies, ensuring only trusted organizations can approve transactions based on specific business logic. In Fabric blockchains, the term "chaincode" refers to smart contracts that have been fully implemented and operational on the network. While there is a distinction between the two terms, they are often used interchangeably. Smart contracts in Fabric can translate business models into blockchain logic using a \ac{sdk}. Fabric \ac{sdk}s provide a comprehensive set of \ac{api} modules that support several general-purpose programming languages, such as Java, JavaScript, and Golang, for developing Fabric-based solutions.  

\subsection{Fabric components}
\subsubsection{Peers}
Fabric peers are software containers that package together the code, tools, libraries, and dependencies necessary for the reliable operation of the blockchain network. They also serve as the physical hosts for ledgers and smart contracts, which are fundamental to any Fabric solution. Peers come equipped with \ac{api}s that enable external applications to access their services. The relationship between peers, smart contracts, and ledgers is very close. Smart contracts, essentially pieces of code, can interact with the ledgers stored on the peers. The primary role of a peer is to host instances of smart contracts and ledgers, providing redundancy for the business logic and data within the network. This redundancy is critical for eliminating single points of failure, which is one of the key advantages of any \ac{dlt}. Peers can manage multiple ledgers and smart contracts simultaneously. However, it is also possible for peers to exist within the network without any installed smart contracts or active ledgers. Depending on the specific requirement, peers can be endorsers, committers, anchors, leaders, and gateways, each having distinct roles, responsibilities, and privileges. 
\begin{itemize}
    \item \textit{Endorsers:} can execute chaincode, validate transactions, and respond to clients.
    \item \textit{Committers:} maintain a copy of the ledger and commit valid transactions to it.
    \item \textit{Anchors:} serve as connection points between different organizations in the network to facilitate communication within a blockchain channel.
    \item \textit{Leaders:} coordinate communication between the ordering service and other peers. They retrieve blocks from orderers and propagate them across the network.
    \item \textit{Gateways:} are the intermediaries between client applications and the Fabric network. They are responsible for submitting transaction proposals from clients to endorsing peers, consolidating endorsements, and delivering endorsed transactions to the ordering service.  
\end{itemize}

\subsubsection{State Database}
Fabric supports two types of state databases for ledger implementation: LevelDB and CouchDB. LevelDB is the default implementation used within peer nodes in Fabric networks. It stores data as key-value pairs and supports basic operations such as key lookups, key range queries, and composite queries. While LevelDB is a simple, lightweight, and convenient option for certain applications, it has limitations when managing large and complex datasets. In contrast, CouchDB is an open-source No-SQL database that operates independently of its peers in Fabric networks. It provides advanced query capabilities based on \ac{json}\footnote{\ac{json} is a data format that uses human-readable text to store and transmit information. It is typically used in web applications using the server-client operation paradigm.} data modeling and chaincode indexing, making it more efficient for querying extensive datasets with complex structures. It is essential that the database type is consistent across all network peers, and the choice must be made before launching the network. Fabric does not support switching between database types due to incompatibility issues. LevelDB uses a binary-based data format, while CouchDB relies on a JSON-based data model. Regardless of the chosen type, a state database maintains two crucial records: the current state of objects in the network (i.e., the world state database) and the history of transactions that led to that state (i.e., the blockchain database). \\

\subsubsection{Organizations}
The organizations in Fabric refer to groups of peers operating under different incentive schemes, which often do not align. Organizations are typically enterprises that own infrastructure in a marketplace and implement their business logic. Fabric provides a trust model that enables organizations with different business models, possibly market competitors, to transact with each other on a transparent and secure platform. In a standard Fabric setup, multiple organizations form a consortium that runs and maintains a blockchain network. Fabric supports decentralized administration by allowing participant organizations to create and manage their networks rather than relying on a single central administration point.\\

\subsubsection{Channels}
When multiple organizations and their peers come together to interact and trade assets in a Fabric network, they operate within the constraints of a blockchain channel communication space. In Fabric terminology, a channel is a mechanism that allows a specific group of peers to communicate and interact privately. Within a channel, these peers agree to run and maintain a distributed ledger governed by rules and transaction logic defined by smart contracts, both specific to that channel. Not all peers within an organization can or may join the same channel, although it is common for all peers to belong to at least one channel. We can think of channels as private blockchains operating within a larger decentralized ecosystem. Fabric can be viewed as a collection of multiple independent blockchain channels working together to support the logic of a distributed solution. \\

\subsubsection{Chaincode}
The term chaincode is an inside term coined by the Hyperledger Fabric project to refer to smart contracts Fabric networks. In this context, chaincode consists of immutable software containing the business logic of the blockchain solution. It manages the lifecycle of network objects and establishes a set of common terms, definitions, rules, processes, and data that network participants agree upon before beginning transactions with one another. Each piece of chaincode has an associated endorsement policy, which specifies the trusted organizations that must sign transaction proposals before they are considered valid. The chaincode model in Fabric supports the development of a wide variety of use cases using general-purpose programming languages like Java, Javascript, and Golang. Once developed, a chaincode becomes available in a blockchain channel through a process that includes packaging, installation, approval, commitment, and initialization on all channel peers. Chaincode is central to Fabric's architecture and is typically the primary focus for solution architects and developers. \\

\subsubsection{Client Applications}
These programs utilize \ac{sdk}s and \ac{api}s to connect with peers in a blockchain channel and execute chaincode transactions, resulting in changes to the channel ledger. Client applications serve as a gateway for the interactions between the external world and Fabric blockchains. \ac{sdk}s enable these client applications to connect to peers, generate transactions, and submit transaction proposals to the network. Additionally, they allow them to receive notifications when transactions are ordered, validated, and committed to a channel. Typically, clients are designed to serve a single organization within a channel, but they can also connect with peers in other organizations across the network. The Fabric platform supports \ac{sdk}s for developing client applications in three general-purpose programming languages: Java, Javascript, and Golang. \\

\subsubsection{Ordering Service}
The Ordering Service module consists of a set of designated peers in the network known as ordering peers, responsible for the consensus mechanism within the system. Ordering peers check transactions and organize them in cryptographic blocks, effectively agreeing on the validity and sequence of transactions in the ledger. The ordering peers are orchestrated within a cluster with a leader-follower structure using the Raft algorithm. Raft is designed to prevent single points of failure, meaning that if the leader peer within the cluster crashes, a new leader is elected using Raft's functionality. This feature makes the ordering service logically centralized but physically distributed across the network. The fault-tolerant nature of the ordering service ensures the availability and consistency of consensus services in Fabric. Unlike public blockchains, Fabric employs deterministic consensus. Since the ordering service is always managed by a single peer at any given time, ordered transactions are guaranteed to be unique when added to the ledger. The responsibilities of the ordering service are distinct from those of chaincode execution and endorsement processes that occur at the regular peer level. This separation of duties helps prevent bottlenecks in the network, contributing to the system's overall performance and scalability. \\

\subsubsection{Membership Service Provider}
The \ac{msp} is a vital part of the Fabric framework, as it is responsible for identity validation within the network. The \ac{msp} not only verifies identities but also manages roles within the blockchain. Its default implementation follows a traditional PKI model, using X.509 certificates issued by trusted CAs to serve as digital identities in the network. Organizations within a Fabric consortium can customize their \ac{msp}s in standard configurations. Fabric can specify configuration parameters such as the \ac{msp} identifier, root trust CA, intermediary CAs, and local paths to cryptographic materials in the channel's genesis block during the network's initial setup. The \ac{msp} establishes a record of trust in the blockchain to authenticate who is permitted to participate. Furthermore, the \ac{msp} can assign accepted identities to specific roles, granting them predetermined permissions to perform actions at the organization, peer and channel levels.\\

\subsubsection{Management tools}
The Hyperledger Fabric project offers management tools for administering and maintaining Fabric networks. Hyperledger Cello is an application that provides a visual interface to simplify the creation and management of Fabric ecosystems. Cello operates on a container-based cluster, often integrated into commercially available blockchain-as-a-service platforms. Hyperledger Explorer is a web application designed to monitor performance metrics within Fabric networks. It overviews network components, including organizations, peers, and chaincode. Explorer lets users view blocks, query data, and transactions submitted to the blockchain. Finally, Hyperledger Composer is a modeling framework that integrates existing business logic with Fabric blockchains. This framework enables network architects to design fabric-based solutions more efficiently, offering a user-friendly toolset that accelerates their design and delivery.\\

\subsection{Assembling the Fabric Puzzle}
\begin{figure*}
    \centering
    \includegraphics[width=0.8
\textwidth]{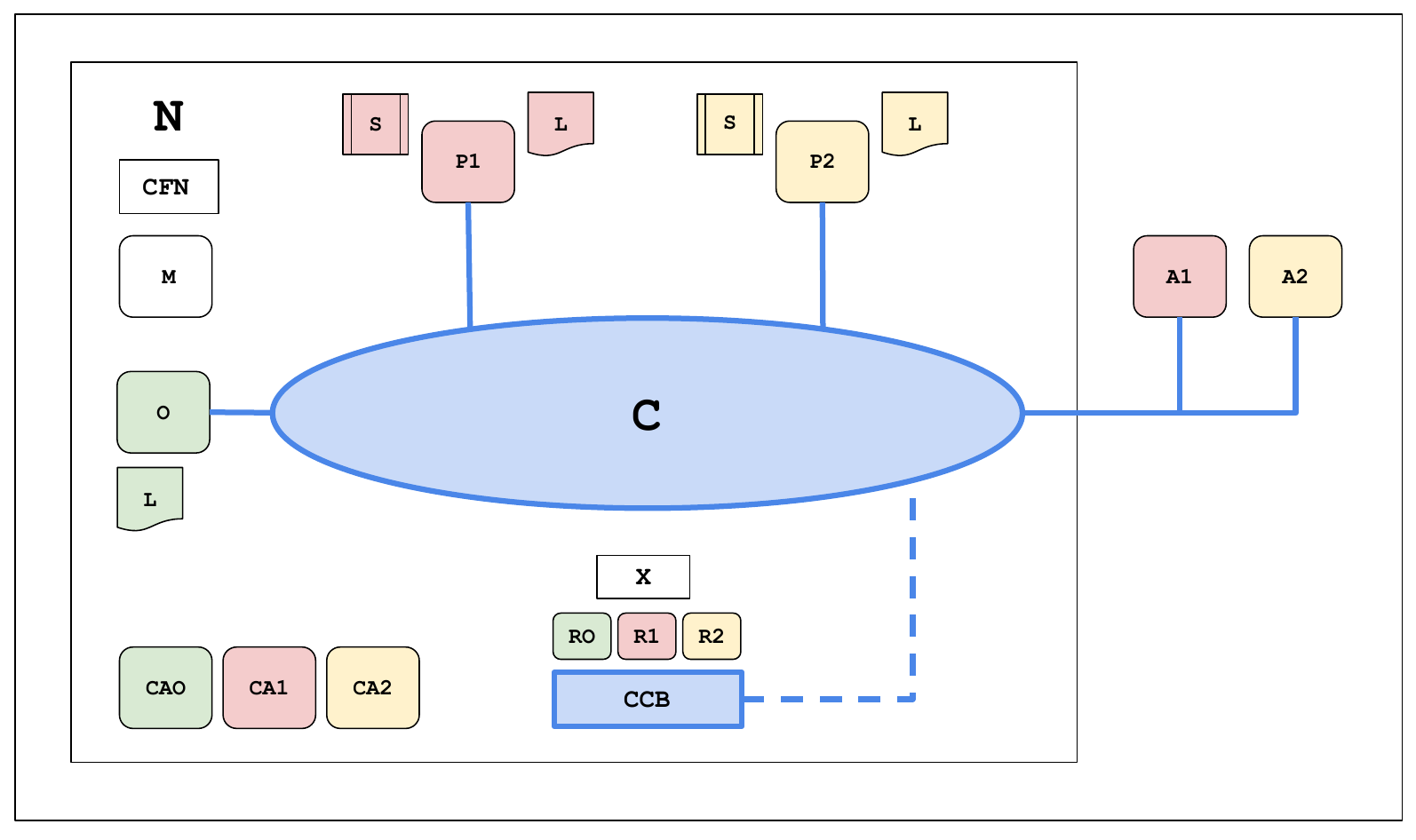}
    \caption{Illustration of a fully formed Fabric network with its main components in action}
    \label{fig:fabricComponents}
\end{figure*}
This section will examine how the components of Hyperledger Fabric form a blockchain network. We will consider a scenario where three organizations, \textbf{\textit{RO, R1}} and \textbf{\textit{R2}}, have agreed to form the consortium \textbf{\textit{X}} to collaborate on a new Hyperledger Fabric blockchain network, referred to as \textbf{\textit{N}}. Organization \textbf{\textit{RO}} will contribute the ordering peer \textbf{\textit{O}}, while  organizations \textbf{\textit{R1}} and \textbf{\textit{R2}} will contribute peers \textbf{\text{P1}} and \textbf{\text{P2}}, respectively. The ordering service servers as the initial administrative point of the network and should be included in every proposed consortium. The identities of the organizations and their peers are validated by certificate authorities \textbf{\textit{CAO, CA1}} and \textbf{\textit{CA2}}. The configuration and policies of the consortium are established in advance by the organizations and documented in the network configuration file known as \textbf{\textit{CFN}}. Once these steps are complete, the organizations \textbf{\textit{RO, R1}} and \textbf{\textit{R2}} will create an application channel \textbf{\textit{C}}, allowing participant peers to interact with one another through transaction proposals. The configuration for channel \textbf{\textit{C}} is agreed upon by \textbf{\textit{RO, R1}} and \textbf{\textit{R2}} and recorded in the channel configuration block, referred to as \textbf{\textit{CCB}}. At this stage, channel  \textbf{\textit{C}}  can be considered created simply by the existence of \textbf{\textit{CCB}} although no peers have joined it yet. Once \textbf{\textit{CCB}} is ready, peers \textbf{\textit{PO, P1}} and \textbf{\textit{P2}} can join \textbf{\textit{C}}. When a peer joins a channel, it automatically receives a copy of the channel ledger (i.e., world state database and blockchain database), holding the current state and history of the network. Specifically, when \textbf{\textit{PO, P1}} and \textbf{\textit{P2}} join \textbf{\textit{C}}, they receive a copy of \textbf{\textit{L}}, the ledger associated to \textbf{\textit{C}}. It is important to note that there are no restrictions on peers joining multiple channels. In production scenarios, it is common for peers to participate in multiple channels and maintain copies of different ledgers. After the peers have joined channel \textbf{\textit{C}}, the chaincode \textbf{\textit{S}} (i.e., smart contract) can be packaged, installed, approved, committed, and initialized across the organizations and peers within network \textbf{\textit{N}}. While \textbf{\textit{S}} is physically installed on the peers, it is logically hosted on \textbf{\textit{C}}. Once \textbf{\textit{S}} is deployed and ready for use, client applications \textbf{\textit{A1}} and \textbf{\textit{A2}} hosted by \textbf{\textit{R1}} and \textbf{\textit{R2}}, respectively, can interact with the Fabric ecosystem. Although \textbf{\textit{A1}} and \textbf{\textit{A2}} are technically not part of the blockchain network, they possess recognized identities issued by \textbf{\textit{CA1}} and \textbf{\textit{CA2}}, allowing them to invoke transactions proposals for querying or updating \textbf{\textit{L}}. Finally, the management tool \textbf{\text{M}} (e.g., Hyperledger Explorer and Hyperledger Cello) is installed on top of the Fabric network for administration and maintenance purposes. An illustration of the fully formed network explained in this section is shown in Figure~\ref{fig:fabricComponents}.

\section{Implementation Blueprint for Hyperledger Fabric Networks}
\subsection{Network Setup}
The first step in launching a Fabric network is to ensure that the physical infrastructure, specifically the computing hardware, is properly connected through a shared communication medium, such as air, fiber optics, or copper. After verifying the physical connections, the necessary software prerequisites must be installed on the computing hardware across the entire blockchain infrastructure. Hyperledger Fabric is a container-based platform that utilizes binary files\footnote{The Fabric binaries are the platform-specific executable files that allow Fabric's launch and proper run.}, package managers,  programming languages, and management tools to operate the blockchain network. Table \ref{fabricPreReqs} shows a complete list of these prerequisites. \\
\begin{table*}[]
\footnotesize
\centering
\resizebox{0.6\linewidth}{!}{
\begin{tabular}{| p{3.3cm} | p{6.5cm} |} 
\hline
\textbf{Software Prerequisites} & \textbf{Description} \\
\hline
& \\
\textbf{Operating System:} & \\
Linux Ubuntu & Unix-based operating system\\
Curl & URL syntax-based software tool for data transmission\\
Node JS & JavaScript runtime environment \\
NPM & Package manager for Node.js\\
& \\
\hline
& \\
\textbf{Programming languages:} & \\
Python & General purpose programming language\\
Java & General-purpose programming language\\
Go & General-purpose programming language\\
Node.js & General-purpose programming language\\
& \\
\hline
& \\
\textbf{Container orchestration:} & \\
Docker & Container-unit software tool for OS-level virtualization\\
Docker Compose & Multi-container software tool for Docker applications\\
Docker Swarm & Container orchestration tool for cluster networks\\
& \\
\hline
& \\
\textbf{Blockchain:} & \\
Fabric binaries & Hyperledger Fabric executable files\\
& \\
\hline
\end{tabular}
}
\caption{Software prerequisites for Fabric's network implementation}
\label{fabricPreReqs}
\end{table*}

\begin{figure*}
    \centering
    \includegraphics[width=0.7\linewidth]{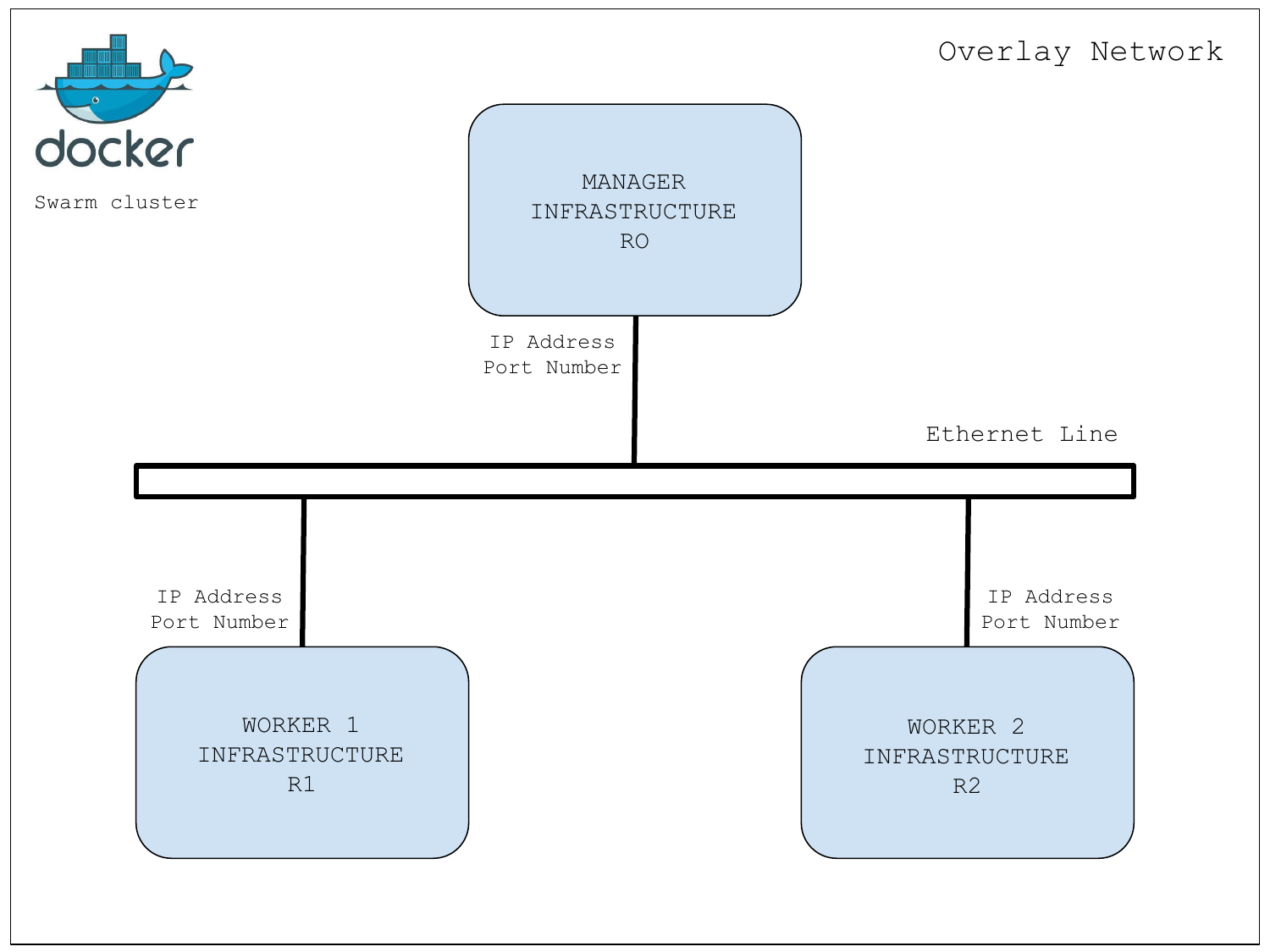}
    \caption{Fabric physical infrastructure.}
    \label{fig:physicalInfrastructure}
\end{figure*}

Once all the required software is installed on the physical infrastructure, the Fabric network will be ready for the initial launch. Typically, the computing hardware, often servers, can be logically organized into clusters using network orchestration software\footnote{A network orchestration software is a tool that helps automate and manage network devices and configurations.}. Docker Swarm is a management tool that facilitates the deployment of container-based networks across multiple physical or virtual hosts. This tool is an effective option for managing physical infrastructure from a single administrative point. In a Docker Swarm cluster, one of the servers is designated as the cluster manager, while the others function as followers or workers. The manager node serves as the central administration point for the entire physical network, from which all Fabric components can be deployed. Figure \ref{fig:physicalInfrastructure} illustrates an example of a Docker-Swarm infrastructure consisting of three physical servers connected via Ethernet line, with one server designated as the swarm manager.

\subsection{Fabric Containers Spin-up}
Once the physical infrastructure is established and organized in a Docker-Swarm cluster, the container-based components of Fabric can be deployed onto the physical network. Let's consider a finite set of independent organizations that form a consortium to launch a blockchain network. Each organization will use computing hardware to host a set of Docker containers with the Fabric components, as specified in the Fabric configuration file (e.g., \textit{configtx.yaml}). Typically, each organization will include a Certificate Authority (CA) to manage identities, peers that host the chaincode responsible for the consortium's logic, and database instances (e.g., CouchDB or LevelDB) to maintain copies of the world state and blockchain. Additionally, each organization will have a \ac{cli} for interaction with external systems. The consortium will also include ordering peers in charge of consensus and membership service to validate identities within the network. \\

\begin{figure}
    \centering
    \includegraphics[width=\linewidth]{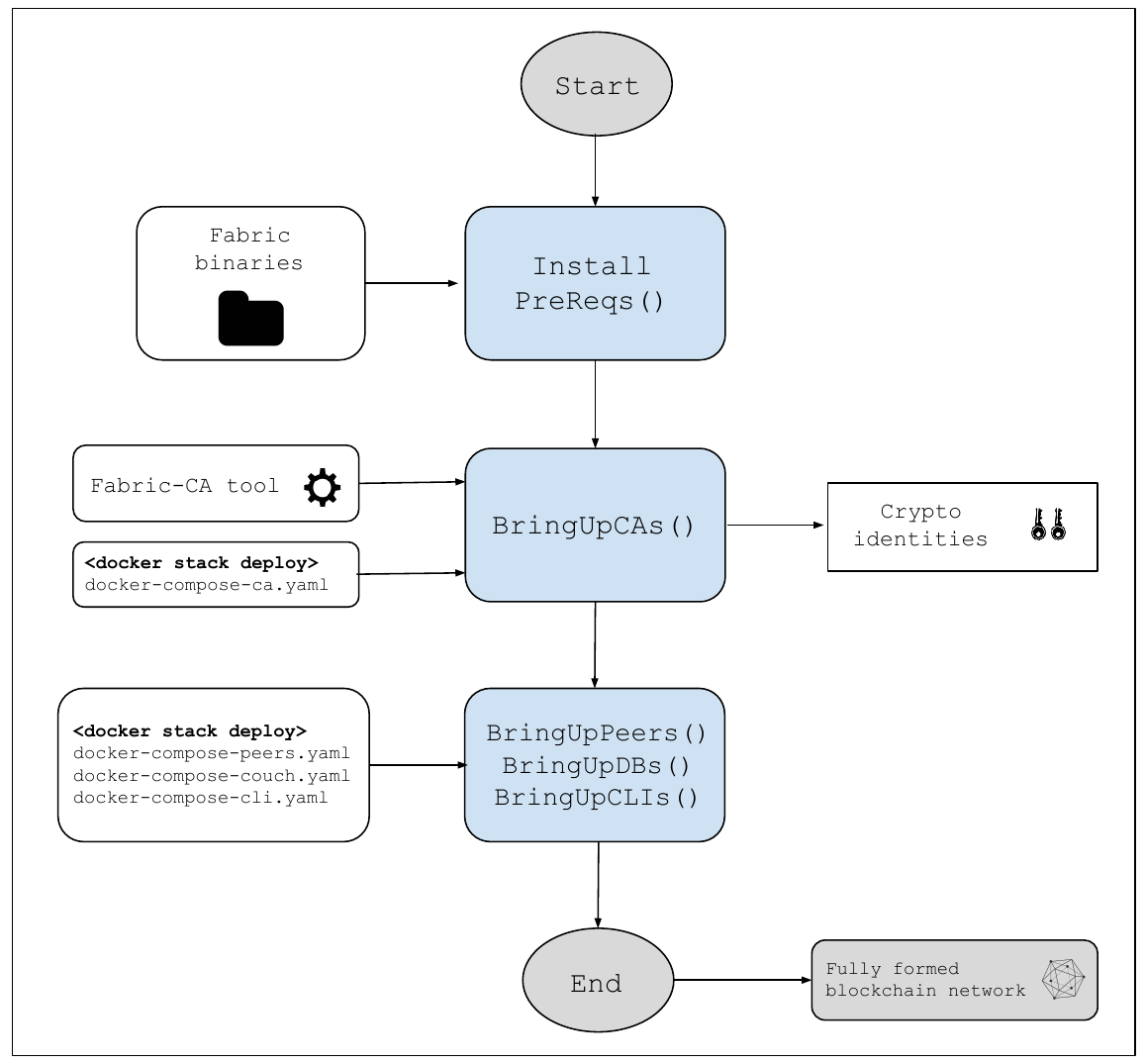}
    \caption{Standard routine to deploy Fabric containers}
    \label{fig:ContainerDeploymentRoutine}
\end{figure}

By default, Fabric provides a \ac{ca} tool called \textit{Fabric-CA}, which operates similarly to a PKI. While this tool is beneficial for prototype development and testing, it is essential to emphasize that production networks should source identity services from trusted, commercially available third-party CAs in the IT sector. \textit{Fabric-CA} deploys CA containers and generates cryptographic identities for network participants, including organizations, peers, and clients, through a configurable process known as \textit{enrollment}. Instructions for enrollment are provided as \textit{Bash} commands included in the tool and can be executed from each organization \ac{cli}. These cryptographic identities enable peers and client applications to sign all their interactions with the blockchain network. Once the CAs and network identities are established, the remaining network components are deployed from the manager node in the Docker-Swarm cluster using the family of \textit{$<$docker stack deploy$>$} commands, executed from the manager's \ac{cli}.  These deployment commands use \textit{YAML}\footnote{YAML is a human-readable data serialization language commonly used for data transmission applications.} configuration files, which outline the container profiles needed for the spin-up process. If container spin-up is successful, ordering peers, organization peers, database instances, and \ac{cli}s will be deployed across the network's physical nodes. For this example, we'll assume that the Fabric network in Figure \ref{fig:fabricComponents} runs on the computing hardware infrastructure in Figure \ref{fig:physicalInfrastructure}. The details of the Docker containers for the example network are provided in Table \ref{tab:ConsortiumComponents}, and the standard deployment routine for Fabric containers is shown in Figure~\ref{fig:ContainerDeploymentRoutine}.

\subsection{Blockchain Channel Creation}
\begin{figure}
    \centering
    \includegraphics[width=1\linewidth]{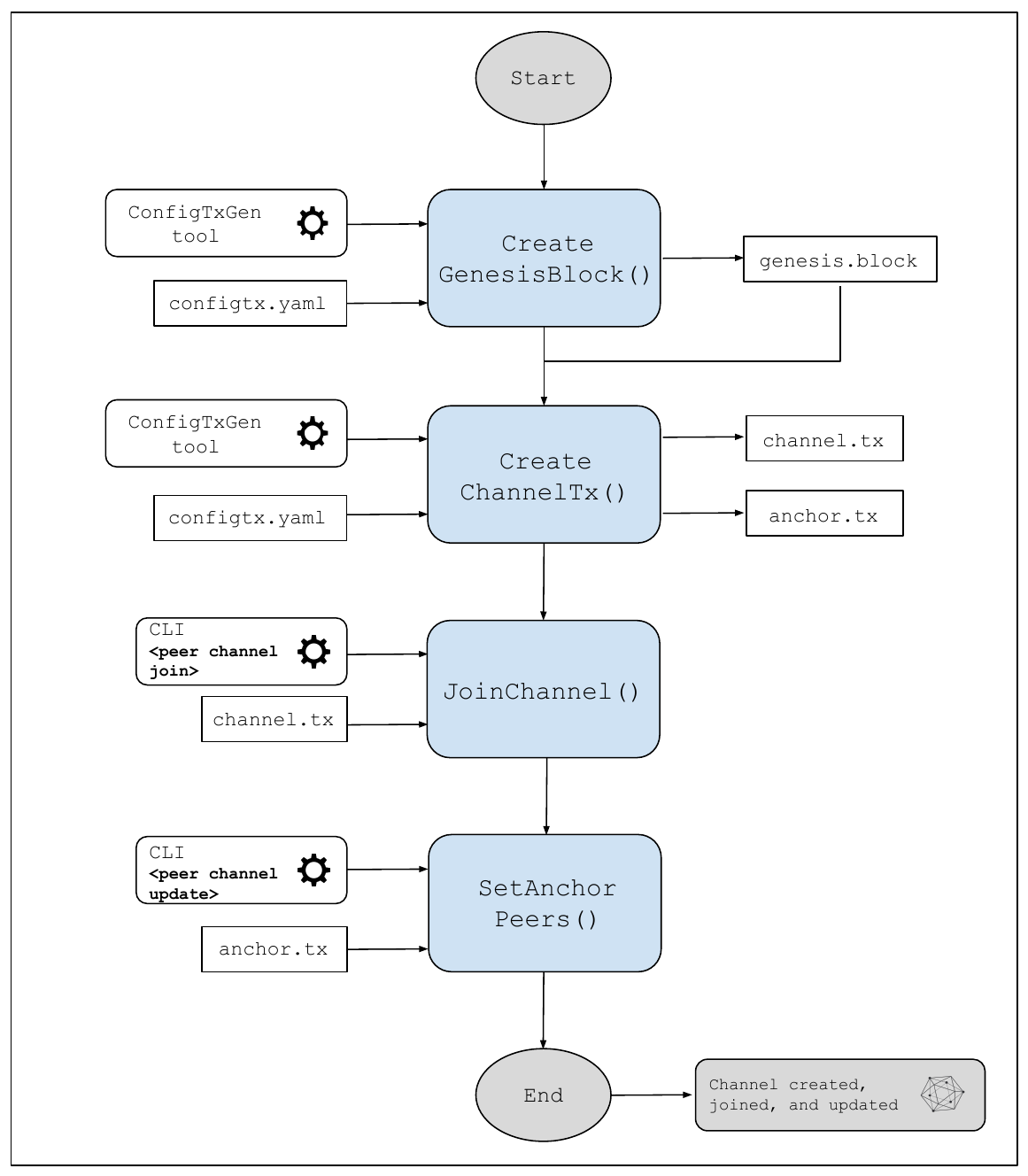}
    \caption{Routine for channel creation}
    \label{fig:channelCreationRoutine}
\end{figure}

\begin{table*}[]
\footnotesize
\centering
\resizebox{0.7\linewidth}{!}{
\begin{tabular}{ | p{2.5cm} | p{2cm} | p{2cm} | p{3.2cm} |} 
\hline
\textbf{Component} & \textbf{Container Name} & \textbf{Physical Node} & \textbf{Network Alias} \\
\hline
& & & \\
\textbf{RO:} & & & \\
Orderer CA & CA0.RO & Manager & CAO.RO.consortium.com\\
Orderer Peer & O.RO & Manager & O.RO.consortium.com \\
Peer Ledger & L.O.RO & Manager & L.O.RO.consortium.com \\
Management Tool & M.RO & Manager & M.RO.consortium.com\\
 & & & \\
 \hline
 & & & \\
\textbf{R1:} & & & \\
Organization CA & CA1.R1 & Worker1 & CA1.R1.consortium.com \\
Organization Peer & P1.R1 & Worker1 & P1.R1.consortium.com\\
Peer Ledger & L.P1.R1 & Worker1 & L.P1.R1.consortium.com \\
Organization CLI & CLI.R1 & Worker1 & CLI.R1.consortium.com \\
& & & \\
\hline
 & & & \\
\textbf{R2:} & & & \\
Organization CA & CA2.R2 & Worker2 & CA2.R2.consortium.com \\
Organization Peer & P1.R2 & Worker2 & P1.R2.consortium.com\\
Peer Ledger & L.P1.R2 & Worker2 & L.P1.R2.consortium.com \\
Organization CLI & CLI.R2 & Worker2 & CLI.R2.consortium.com \\
& & & \\
\hline
\end{tabular}
}
\caption{Hyperledger Fabric components}
\label{tab:ConsortiumComponents}
\end{table*}
After deploying Fabric components in containers, the network is ready to create blockchain channels. Hyperledger Fabric provides the \textit{ConfigTxGen} tool to assist with this process, which is used for creating and joining blockchain channels. This tool generates channel configuration blocks, known as genesis blocks, along with all related channel artifacts, specifically \textit{channel.tx} and \textit{anchor.tx}. To create a channel, \textit{ConfigTxGen} requires a network configuration file (e.g.,  \textit{configtx.yaml}) as input and produces the \textit{channel.tx} and \textit{anchor.tx} artifacts as output. The configuration file contains the identities of the organizations, peers, client applications authorized to participate in the channel, and the information about the anchor peers. When peers want to join the channel, each joining peer must submit a join transaction using the \textit{channel.tx} artifact. The peer becomes a blockchain node if there is a consensus in the channel to approve the transaction. By default, not all peers in an organization can communicate with peers from other organizations within the consortium. Specific peers, known as anchor peers, are designated as contact points for communication across organizations. Hyperledger Fabric utilizes a gossip protocol for peer-to-peer communication, and anchor peers play a crucial role in this protocol by propagating blockchain blocks and transactions beyond their organization's boundaries. To become an anchor peer, a peer must submit an update transaction using the anchor artifact \textit{artifact.tx}. The peer is granted anchor status only if the channel approves this transaction according to the rules outlined in  \textit{configtx.yaml}. Figure~\ref{fig:channelCreationRoutine} illustrates the standard routine for channel creation. \\ 
\subsection{Chaincode Deployment}
As defined in section III, chaincode within the Hyperledger Fabric framework is immutable software that defines network objects (i.e., assets) and provides instructions on how to modify them in the network's ledger. An asset can represent anything depending on the specific logic of the blockchain and is typically structured as an entity consisting of attributes, data, and methods. In chaincode and client applications, assets are usually represented using generic data structures like JSON. Developers can code the blockchain logic within chaincode pieces using general-purpose programming languages such as Java, Go, and Node.js, along with \ac{api}s. Fabric \ac{api}s offer various classes, methods, and interfaces to facilitate the coding of contracts and transactions and to perform operations on the ledger. Chaincode represents the programmable logic of the Fabric network and governs how network assents can be modified through transaction execution. The typical lifecycle of a chaincode consists of the following stages: package, install, instantiate, and upgrade when necessary. Once a blockchain channel is fully operational, chaincode pieces should be installed on the channel endorsers and subsequently approved by the network organizations. It's essential to understand that Fabric endorsers are the network peers responsible for receiving transaction proposals from clients, executing them using their installed instance of the chaincode to verify outputs, and generating endorsements for transactions that align with the expected network logic. Endorsements are required to validate requests for modifying the attributes and data of network assets in the blockchain by a client application. Only transaction proposals that receive sufficient endorsements to meet the network's endorsement policy can be sent to the ordering service for final approval and commitment to the channel's blockchain. Chaincode pieces are generally available per-channel, and a set of access policies defined in the network configuration determine which peers and clients can access and interact with their logic. \\
\subsection{Client Integration}
\begin{figure*}
    \centering
    \includegraphics[width=0.8\linewidth]{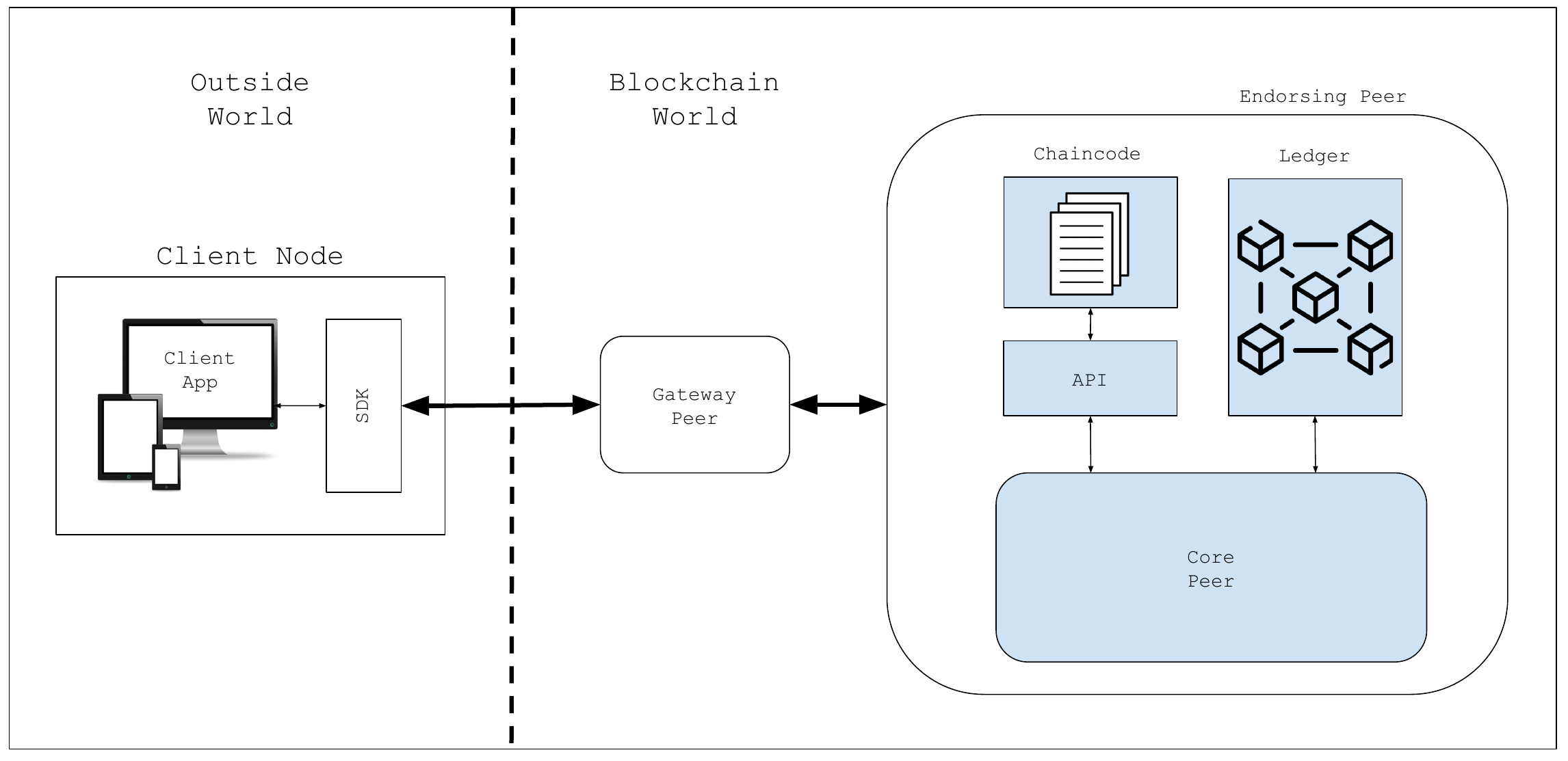}
    \caption{Communication between client applications and channel endorsers through gateway peers.}
    \label{fig:clientIntegration}
\end{figure*}

Client applications are essential for any Fabric solution, as they bridge the outside world and the blockchain infrastructure. Typically written in general-purpose programming languages, these applications connect to the blockchain platform using Fabric \ac{sdk} modules to communicate with peers in a blockchain channel. Clients connect with gateway peers within their parent organization to submit transaction proposals to the network. The configuration details for gateway peers can be found in the connection profile created by parent organizations before joining the network consortium. Gateway peers are responsible for routing transaction proposals to endorsing peers in blockchain channels. After transaction proposals reach endorsers, these peers utilize their local copy of the chaincode and a Fabric \ac{api} to simulate, approve, or reject proposals submitted to the network. When endorsements (i.e., transaction approvals) or rejections are ready, endorsing peers send their decision back to the clients through gateway peers. Then, clients collect responses from all the channel endorsers and check if the proposal complies with the endorsing policy of the blockchain channel. If so, proposals are sent to the ordering service for transaction aggregation, ordering, block creation, and final committing to the channel distributed ledger. An illustration of how clients communicate with endorsing peers can be found in Figure~\ref{fig:clientIntegration}.
%
\subsection{Overview of the Communication Model}
\begin{figure*}[h]
    \centering
    \includegraphics[width=0.65\textwidth]{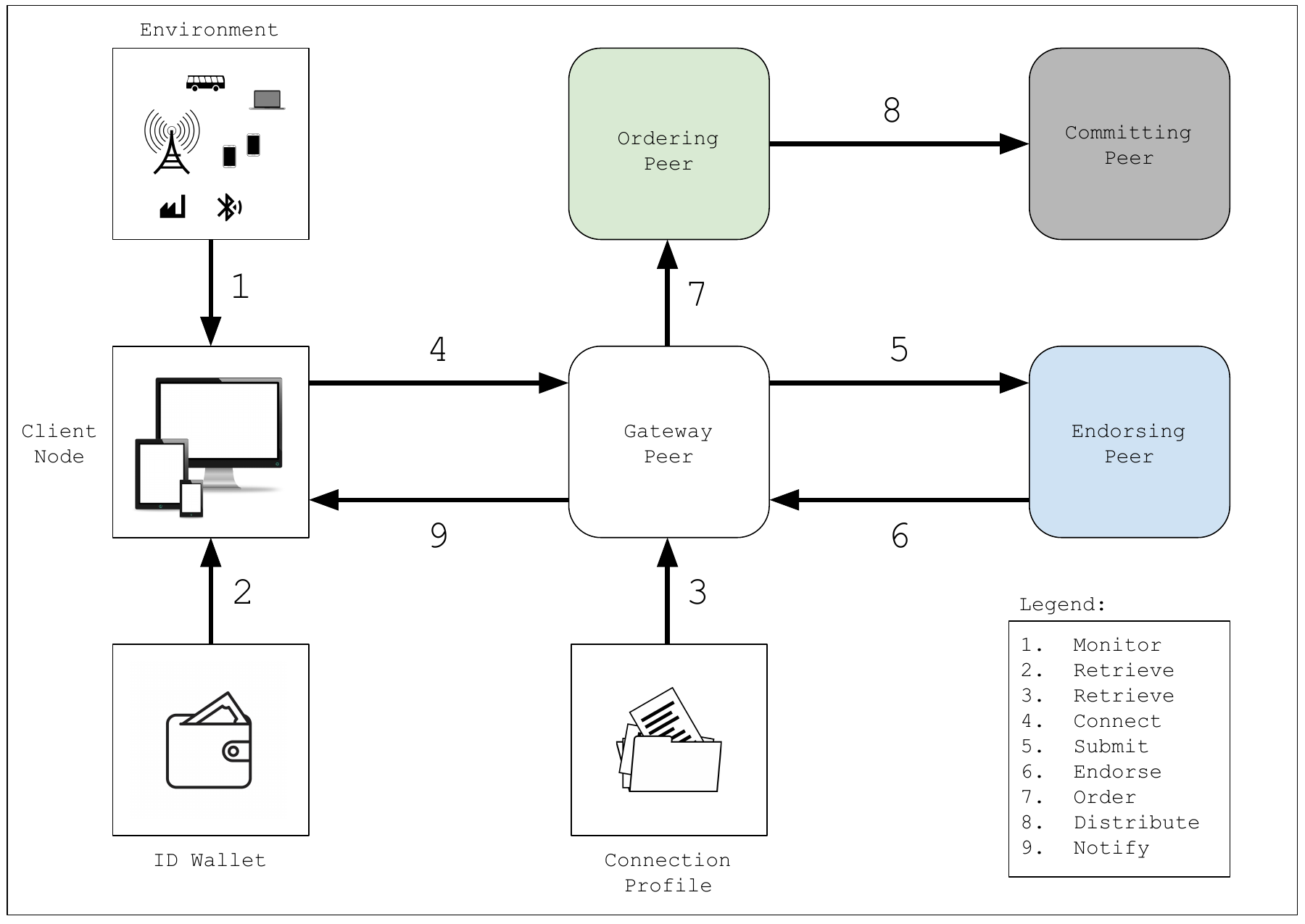}
    \caption{Fabric communication model}
    \label{fig:communicationModel}
\end{figure*}

Once all components of the Fabric platform are fully integrated, the blockchain network is ready for production. Normally, transaction proposals are initiated by changes in the environment of blockchain networks (1). Typically, client nodes detect those changes and can submit transactions by retrieving \textit{x.509} digital certificates from their ID wallets (2). These identities correspond to pre-authorized users who have the clearance to access the network. Next, they load the connection profile of designated gateway peers within their parent organizations (3). Both identity wallets and connection profiles are stored on the client's local file systems. Gateway peers provide entry points for clients to access a blockchain channel and propagate transaction proposals to endorsing peers within this blockchain ecosystem (4). From gateways, transaction proposals are routed to designated endorsing peers in the channel to set off the consensus mechanism (5). Depending on the validity of these transaction proposals, endorsers may approve, sign, and return them to clients (6). The clients then collect all received endorsements and send the endorsed transactions to the leading ordering peer through the gateway peer again (7). The ordering peer then checks the endorsed transactions against the endorsement policy governing the channel. If transaction proposals comply with the endorsement policy, the ordering service generates blocks with verified transactions and distributes them to committing peers (8), which then proceed to add these blocks to their local copies of the ledger. Finally, when blocks are added to committing ledgers, client applications receive a notification alerting them of the end of the transaction lifecycle. A summary of the communication model involving the main components of the Fabric network is provided next and illustrated in Figure \ref{fig:communicationModel}.
\begin{enumerate}
    \item A client node can monitor and detect changes in the environment of a blockchain solution.   
    \item The client initiates a transaction proposal by retrieving its \textit{x.509} identity from the local ID wallet.
    \item The client also retrieves the connection profile of the designated gateway peer in its parent organization. 
    \item The client can access a blockchain channel using the gateway peer as the blockchain entry point to propagate the transaction proposal. 
    \item From the gateway peer, the transaction proposal is routed to the endorsing peers in the channel to trigger the consensus mechanism.
    \item Endorsing peers verify the proposal's validity by using their local copy of the chaincode and may endorse (i.e., approve) or refuse it depending on its validity. 
    \item The client node collects endorsements and sends the endorsed proposal to the ordering service. 
    \item The ordering service verifies the endorsed transaction against the endorsement policy and packs it, along with other verified transactions, in a data block that is later distributed to committing peers. 
    \item When committing peers add the recently created block to their local copies of the ledger, the client node receives a notification alerting the end of the transaction lifecycle.
\end{enumerate}
\section{Key Takeaways}
Trust models are essential in any network to verify the integrity of its participants and establish a framework for the overall reliability and resilience of the system. Trust models can be centralized or decentralized, each offering unique benefits and challenges. Distributed trust models refer to network models involving collective management of integrity mechanisms by many network participants, ensuring no single entity controls the network's rules and decisions. \\

A notable example of distributed trust models is blockchain, which refers to peer-to-peer networks that utilize advanced cryptographic techniques and consensus algorithms to maintain a distributed database of records (i.e., the ledger) among independent nodes that do not trust each other. By integrating these elements, blockchain networks enforce confidence among participants. Distributed consensus functions like a digital "democracy," allowing network participants to collectively make decisions within the network without the oversight of a central authority with concentrated power. By maintaining a local copy of the ledger, blockchain participants act as guardians of the network's historical records, providing transparency and accountability to the system and thereby reinforcing trust. \\

Consortium blockchains represent a specific type of blockchain in which governance, trust, and control functions are shared among several participant organizations, each with potentially different incentives. Typically, the architecture of consortium blockchains offers open, modular, and lightweight decentralization services. Hyperledger Fabric is a consortium blockchain platform that has gained traction across various economic sectors, including healthcare, finance, supply change, and engineering. Fabric provides membership, consensus, ledger, and smart contract services for modular and versatile decentralized solutions that are relatively straightforward to deploy using general-purpose programming languages, container orchestration software, and the openly available Fabric source code. 
\section*{Acknowledgment: Why are We Interested in Blockchain Technologies?}
In the telecommunications field, trust models are essential for orchestrating computer networks such as mobile wireless systems. As mobile systems grow larger with every generation, network management models based on centralized trust are increasingly problematic, leading to issues like high latencies, backhaul congestion, and elevated risk of cyber attacks due to the increased presence of potentially malicious actors within the network. From the trust model perspective, consortium blockchain technologies have the potential to address these challenges by enabling fast and lightweight platforms offering membership, consensus, ledger, and smart contract services that support distributed, secure, automated, and efficient infrastructure for communication systems. \\

This draft is a compilation of blockchain notes and annotations collected in the context of a research project on integrating blockchain technologies into the architecture of next-generation communication systems. Specifically, our work explored the possibility of combining the Hyperledger Fabric blockchain platform as a distributed trust model for a resource-sharing architecture for computing servers placed at the edge of a communications network. Our blockchain model enables a truthful collaboration scheme for edge servers to share computing tasks with the rest of the network. The model ensures secure and private server collaboration using the blockchain services offered by the Fabric platform. These services include membership, consensus, ledger, and smart contract solutions. Our work culminated with the implementation of a proof of concept demo for our blockchain-based task-sharing architecture that required running the Hypeprledger Fabric blockchain on a set of physically interconnected PCs, simulating the edge layer of a communication network. \\

The project was sponsored by the University of Manitoba, and it was part of the research portfolio of the Wireless Communications, Networks, and Services (Wicons) laboratory led by Dr. Ekram Hossain. The details and results of our research effort can be tracked in the following publications: \cite{06} \cite{07} \cite{08} \cite{09} and \cite{10}.
\bibliographystyle{IEEEtran}

\begin{thebibliography}{2}
\bibitem{01} S. Nakamoto, “Bitcoin: A peer-to-peer electronic cash system,” https://bitcoin.org/bitcoin.pdf, 2008.
\bibitem{02} The Linux Foundation Project \& The Hyperledger White Paper Working Group, “Introduction to hyperledger fabric white paper,” 2018.
\bibitem{03}The Linux Foundation Project \& The Hyperledger Performance and Scale Working Group,“ White Paper: Hyperledger blockchain performance metrics,” 2018.
\bibitem{04} IBM, “Hyperledger fabric: A distributed operating system for permissioned blockchains,” Technical Report, 2018.
\bibitem{05} S. Mazumdar and S. Ruj, “Design of anonymous endorsement system in hyperledger fabric,” IEEE Transactions on Emerging Topics in Computing, pp. 1–1.
\bibitem{06} A. Vera-Rivera, A. Refaey, and E. Hossain, “A blockchain framework for secure task sharing in multi-access edge computing,” IEEE Network, pp. 1–8, 2020.
\bibitem{07} A. Vera-Rivera, A. Refaey, and E. Hossain, “Task sharing and scheduling for edge computing servers using hyperledger fabric blockchain,” IEEE GLOBECOM Global Communications Conference, Edge Computing Workshop, 2021.
\bibitem{08} A. Vera-Rivera, A. Refaey, and E. Hossain, “Blockchain-based collaborative task offloading in mec: A hyperledger fabric framework,” IEEE ICC International Conference on Communications, 6G Workshop, pp. 1–8, 2021.
\bibitem{09} A. Vera-Rivera, E. Hossain, and A. Refaey, “Exploring the intersection of consortium blockchain technologies and multi-access edge computing: Chronicles of a proof of concept demo,” IEEE Open Journal of the Communications Society, vol. 3, pp. 2203–2236, 2022.
\bibitem{10} A. Vera-Rivera, “Design and implementation of a blockchain-based task sharing service for edge computing servers using the hyperledger fabric platform,” Masters thesis, University of Manitoba, Winnipeg, MB. Canada., October 2022, available at https://mspace.lib.umanitoba.ca/items/fcbe71bc-87a5-441f-a798-f097527f2218.
\end{thebibliography}

\printacronyms

\begin{IEEEbiography}[{\includegraphics[width=1.1in,height=1.10in,keepaspectratio]{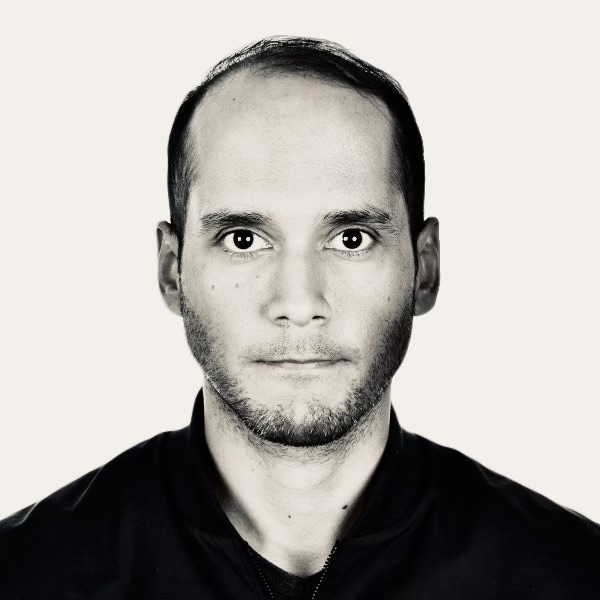}}]
{Angelo Vera-Rivera} is a Research Technician in the Department of Electrical and Computer Engineering at the University of Manitoba. He works at the Wireless Communications, Networks, and Services Laboratory (Wicons Lab) under the supervision of Dr. Ekram Hossain. He holds an M.Sc. degree in Electrical and Computer Engineering from the University of Manitoba (Canada, 2022), an M.Sc. in Telecommunications from George Mason University (United States, 2015), and a B.Sc. in Electronics and Telecommunications from Escuela Superior Politecnica del Litoral (Ecuador, 2011). He is also a member of Engineers Geoscientists Manitoba (EIT). His research interests focus on the intersection of Blockchain and Edge Computing technologies with next-generation communication systems. 
\end{IEEEbiography}
\begin{IEEEbiography}[{\includegraphics[width=1in,height=1.25in,clip,keepaspectratio]{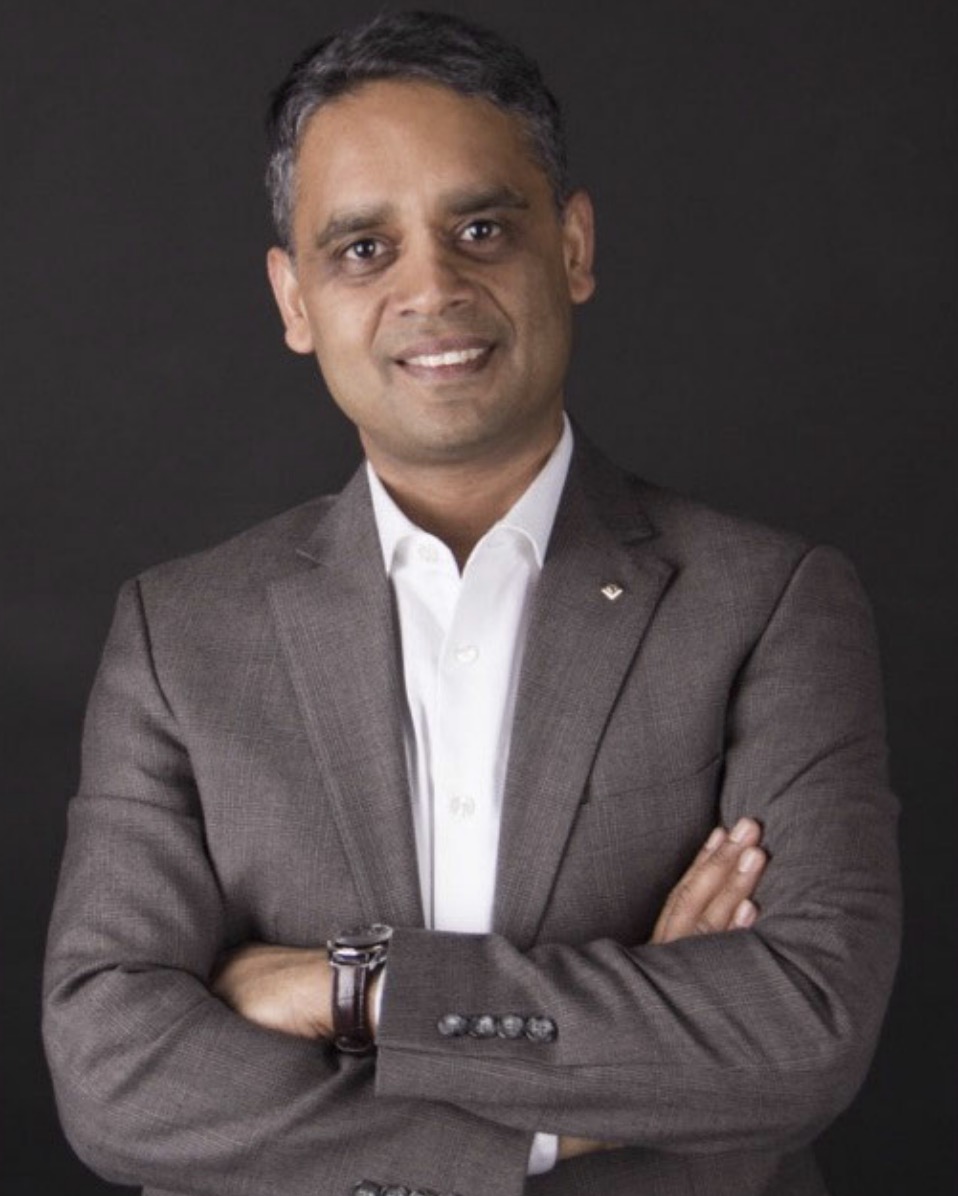}}]{Ekram Hossain}(Fellow, IEEE) is a Professor and the Associate Head (Graduate Studies) of the Department of Electrical and Computer Engineering, University of Manitoba, Canada. He is a Member (Class of 2016) of the College of the Royal Society of Canada. He is also a Fellow of the Canadian Academy of Engineering and the Engineering Institute of Canada. He has won several research awards, including the 2017 IEEE Communications Society Best Survey Paper Award and the 2011 IEEE Communications Society Fred Ellersick Prize Paper Award. He was listed as a Clarivate Analytics Highly Cited Researcher in Computer Science in 2017-2024. Previously, he served as the Editor-in-Chief (EiC) for the IEEE Press (2018–2021) and the IEEE Communications Surveys and Tutorials (2012–2016). He was a Distinguished Lecturer of the IEEE Communications Society and the IEEE Vehicular Technology Society. He served as the Director of Magazines (2020-2021) and the Director of Online Content (2022-2023) for the IEEE Communications Society.
\end{IEEEbiography}

\end{document}